\documentclass{PoS}
\usepackage{graphicx}
\usepackage{epsfig}

\title{GPU computing for 2-d spin systems: CUDA vs OpenGL }

\ShortTitle{GPU computing for 2-d spin systems}

\author{Viola Anselmi\\
        University of Parma\\
        E-mail: \email{viola.anselmi@studenti.unipr.it}}
\author{Giovanni Conti\\
        University of Parma\\
        E-mail: \email{giovanni.conti@studenti.unipr.it}}
\author{\speaker{Francesco Di Renzo}\\
        University of Parma and INFN \\
        E-mail: \email{francesco.direnzo@unipr.it}}

\abstract{In recent years the more and more powerful GPU's available on the PC market have attracted attention
as a cost effective solution for parallel (SIMD) computing. CUDA is a solid evidence of the attention
that the major companies are devoting to the field. CUDA is a hardware and software architecture
developed by Nvidia for computing on the GPU. It qualifies as a friendly alternative to the approach
to GPU computing that has been pioneered in the OpenGL environment. We discuss the application of
both the CUDA and the OpenGL approach to the simulation of 2-d spin systems (XY model).}

\FullConference{The XXVI International Symposium on Lattice Field Theory \\
		 July 14 - 19, 2008\\
		 Williamsburg, Virginia, USA}

\begin{document}

\section{GPGPU: a cost-effective approach to parallel computing}

In recent years a new acronym entered the scene of high performance computing: GPGPU, which stands for General Purpose computations on Graphics Processing Units \cite{GPGPU}. The main point is easily stated. Since an impressive growth is taking place in the field of GPU (Graphics Processing Units) technology, it makes sense to try to exploit GPU performances not only for graphics, but also in other applications. It turns out that graphic processing is a prototype example of parallel SIMD (Single Instruction on Multiple Data) computing, so that every problem that can be mapped to a SIMD implementation is a GPGPU candidate. Does it make sense to try this approach? The bottom line is cost-effectiveness. GPU technology is a typical commodity technology: a huge market (many PC's are actually mainly used for games) results in low prices. In view of this, if a problem can be ported to a GPU environment, then most probably this implementation is going to be very cost-effective. \\
One of the fascinating point in computer science is that many applications actually start almost as DIY (Do It Yourself) and then turn into a common interest for a big community: you can learn the current trends on the web, both from blogs and from specialized web sites. GPGPU is not an exception with this respect (\cite{GPGPUdeut} and \cite{GPGPUweb} are good examples). Scientific applications made their entrance in the field quite early, and big experts are in the lattice community: \cite{Fodor} is a pioneering work, quite well known among experts\footnote{It was \cite{Fodor} that actually triggered the interest that eventually resulted in this work}. \\
An interesting point to make is that by now GPGPU is regarded as a commercial opportunity by big companies: {\tt CUDA} (the Nvidia environment we will refer to) is a very good example.

\subsection{The technological scenario}

Whenever GPGPU has to be introduced, there is a plot which is most often displayed: we are talking of the comparison between the raise in GPU computing capabilities (as measured in GFlops in peak performance) and CPU computing performance. We refer the reader to \cite{CUDA} for an example\footnote{Needless to say, most of these comparisons are actually GPU manufacturers pride.}. Roughly speaking, in recent years GPU's gained an order of magnitude with respect to CPU's. What is important as well, also bandwith increased in a similar way. An important caveat is that all this computing power has up to now been delivered in single precision: while there is apparently no compelling demand for double precision in graphics, an eye to be kept on other applications is driving GPU manufactures attentions on double precision as well. \\
Things are never definitive in the field of computing performances, still it is not so difficult to understand the trend we have just referred to. In the end, every computer is from a logical point of view made of three basic resources: they are usually referred to as control, data-path and memory. From this point of view, every computer design is a given choice in resources allocation. Now, resources of modern CPU's are to a large extent dedicated to control and memory: branch-prediction technologies and big caches are typical examples of this trend. In the end, the component of the processor that perform arithmetic operations (datapath) amounts to a minor fraction of resources. Things are just the other way around in a typical GPU design: the vast majority of resources are put on ALU's (Arithmetic Logic Unit). If one takes into account what is the basic structure of graphic computations this does not come as a surprise. \\
The main issue is that many steps are involved in the process of getting a 2-d bitmap from a 3-d image: the bottom line is that images have to be mapped onto a monitor screen. These steps are typically organized in what is called the \emph{graphic pipeline}. Getting into this subject in details is by far out of reach in the context of this presentation. It is enough to mention the basic stages that are in place. First of all, there is a \emph{modeling} stage: roughly speaking, the image is mapped onto geometric primitives. Then there is a bunch of operations which have to do with geometric elaborations: \emph{viewing} (mapping of the scene onto a plane as seen from a virtual camera), \emph{clipping} (every elements which is not visible should be cut), \emph{lighting} and \emph{shading} (quite obvious meaning). \emph{Texturing} and \emph{rasterization} are the stages at which the image is finally mapped to a bitmap, \emph{i.e.} a 2-d bunch of pixels, each of a definite color. It is important to keep in mind that a modern GPU provides a hardware implementation of this pipeline. As usual in a pipeline, the output of a stage is input for the following one. In any stage of the graphic pipeline, a large amount of data are processed at the same time in a SIMD way. Also, memory latencies can be hidden by computations without the need for big caches. In particular, a very intensive stage of the process is the so-called \emph{fragment shading} stage (which has to do with  texturing): because of this, GPGPU application try to make intensive use of the GPU resources which are allocated to this stage. \\
A last remark on data formats is in order: key data types are \emph{textures}. Basically, they are (2-d) matrices, whose entries are termed \emph{texels}. As already pointed out, this is easy to understand: images
should be eventualy mapped to a bunch of pixels on the screen. Vector types are a natural choice in this context: in particular, {\tt RGBA} \emph{textures} (see figure) account for
colours (Red, Green, Blue) and Alpha channel (transparency). 
\begin{figure}[!htb]
\begin{center}
\includegraphics[scale=0.58,clip=true]{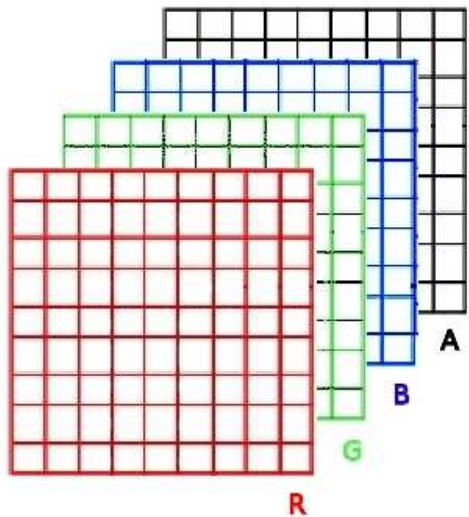}
\end{center}
\vspace{-21mm}
\caption{Pictorial representation of a {\tt RGBA} texture.}
\end{figure}

\subsection{2-d spin systems as a case of study}

We decided to perform a simple benchmark computation by implementing a Hybrid MonteCarlo \cite{HMC} simulation of the XY 2-d spin model, with Hamiltonian
\begin{equation}
	H = - \sum_{<i,j>} \sigma_i \cdot \sigma_j.
\end{equation}
We stress that the model was only taken as a laboratory. We wanted to understand how efficient a GPGPU implementation could be. In the following we will report on two different implementations of the same MC simulation. They are representative of two approaches at our disposal. \\

The first approach to GPGPU has been based on {\tt OpenGL} (a standard graphics library), and so we took this as a first option. 
Basically, one \emph{talks to GPU as if one were performing standard image processing}. In
other words, the computation enters the graphic pipeline. In view of what has already been pointed out, the standard choice is to enter the so-called fragment stage. Both the upload of code and data and the collection of results are a bit funny (again, we are pretending to perform standard graphics). \\
The second implementation was in the framework of {\tt CUDA}. Nvidia (a major GPU manifacturer) provides a hw/sw architecture to actually access the GPU as a (parallel) coprocessor. This approach is relatively novel with respect to pioneering GPGPU applications. Having to devise a numerical project on a GPU now, {\tt CUDA} is the most direct approach to start with and this is why we first investigated it. In view of this, this implementation is also the first we report on in our presentation. \\

Before proceeding to illustrate the peculiar features and the performance results for the two approaches, we pin down some common general features.
\begin{itemize}
\item As for generation of momenta: flat random numbers were generated on CPU,
conversion to gaussian was performed on GPU.
\item Some operations are critical with respect to single/double precision: global sums
of the energy were performed on CPU (as well as Metropolis step). The core of the parallel computation is the leap-frog integration of equations of motion.
\item Results were cross-checked with series expansions in the high temperature
regime \cite{But} and with a reference HMC (this was done also for acceptance). The reference serial code was run on an {\tt Intel Conroe}. In order to get a cheap estimate, the serial code was also taken as a
reference (at fixed HMC parameters) for performance evaluation.
\end{itemize}

\section{The CUDA approach}

We have already referred to the interest major GPU manufacturers have been devoting to GPGPU applications. This does not come as a surprise, as it opens the way to potential commercial opportunities. A good example is provided by Nvidia. The company (one of the leaders in the field) offers both what they call a hw/sw architecture ({\tt CUDA}) enabling GPGPU applications on commodity hardware, both dedicated products (Tesla).

\subsection{An overview}

Nvidia calls {\tt CUDA} \cite{CUDA} a hw/sw architecture which is intended to enable parallel
programming on GPU. The main idea is that no dedicated hardware is requested: the graphic card on your PC has to be regarded as your (parallel) coprocessor. In the end, they distribute a driver to access the device (the list of compatible GPU models extends to last three generations) and a toolkit enabling a programming environment which is basically an extension to C. Users are provided with
\begin{itemize}
\item A \emph{nvcc} C compiler (to code on the device), togheter with debugging and profiling tools.
\item A couple of high level basic scientific libraries (worked-out {\tt CUDA} implementations of FFT and BLAS).
\item Quite extended documentation with a collection of worked-out examples.
\end{itemize}

Notice that in the {\tt CUDA} environment there is no explicit reference to what 
a GPU is mainly devoted to, \emph{e.g.} there is no explicit reference to 
what role each device plays in the above mentioned graphic pipeline. 
Basically, {\tt CUDA} asks the user to regard a recent GPU as a collection of
multiprocessors, each made of several processors. In order to understand 
this point and the following, the reader is referred to the figures in \cite{CUDA}. 
One should be aware of the hierarchy of memories at disposal on the device. In particular, 
there is a \emph{Device Memory} on which any processor can write and from which any processor 
can read. There is on the other hand a much faster \emph{Shared Memory}, which is read/write 
accessible to processors within the same multiprocessor. \\
Basic {\tt CUDA} tool is a driver enabling the access the device in a natural way. In
particular, the language enables the user to upload/download data to/from the device
memories. One is then entitled to make one's own choice on
where data should reside. In the end, resources are limited and all the
game goes back to their allocation. \\
A process running on CPU (host) can start (several) kernels on GPU (device).
Basic organization is thread-based: \emph{blocks of threads}
come in a \emph{grid of blocks}.
As already said, there are commands to upload/download
data to/from the device memory. To execute a kernel the call will be something like 
\begin{verbatim}
My_kernel<<<dimG,dimB>>>(my_arg_1, ... , my_arg_n)
\end{verbatim}
where {\tt dimG} is the dimension of the grid of blocks, each of which is of dimension 
{\tt dimB}. Threads within a block can be synchronized
(they are assigned by the system to the same multiprocessor) and they typically access
shared memory. There are limitations to the number of
threads within a block and of blocks within a
grid. There are of course also limitations imposed by
shared memory dimension. It is clear that there can be no one-to-one correspondence 
in beetwen threads and processors: actual allocation
of resources is up to the system. This is performed by executing one or more blocks on each multiprocessor by \emph{time slicing}.

\subsection{Our implementation}

Allocating the lattice can be really straightforward. Basically, a very direct recipe is 
to map a site to a thread: the grid of blocks is the lattice itself and the blocks of threads are 
sublattices which are taken care of by the same multiprocessor. The lattice is first allocated to 
the \emph{Device Memory}, then sublattices can be moved to \emph{Shared Memory} when threads (sites) 
need to communicate, \emph{i.e.} during the evolution of momenta, when they have to access nearest 
neighbors. To do that one can take advantage fo the standard 
recipe to duplicate sublattices borders which are not updated, but can be accessed in the same way of 
any other site. \\
We have to keep our own balance optimizing the usage of resources. We have to 
admit that this implementation was easy and fast, but, in view of what we learned in the other 
approach, it is obvious that one can do better than we did.

All in all, implementation was very fast, optimization is most probably not
complete, but getting substantial gain was easy. Implementation was performed on a {\tt Nvidia
GeForce 8800GTX}. Performances we will report refer to a version which does not even perform gaussian generation step on the GPU.

\section{The OpenGL approach}

As already said, the pioneering GPGPU applications were in the framework of 
standard graphic libraries, namely {\tt OpenGL}. In particular, usage has 
been made of  
{\tt GLSL} (Shading Language): another (less direct) extension to C, providing an
environment for access to GPU in the OpenGL framework.

\subsection{An overview}

\begin{itemize}
\item Vector variables are a natural choice (\emph{e.g.} {\tt vec2}, {\tt vec3}, {\tt vec4}). 
{\tt vec4} are an obvious choice for {\tt RGBA} textures.
\item Special output variables are present (e.g. vec4 {\tt gl\_FragColor}): always keep in mind you are
supposed to process images.
\item There is a math library available.
\end{itemize}

The basic {\tt GLSL} approach to GPGPU is quite easy to explain: your code will be 
essentially devised to enter the graphic pipeline. As already stated: \emph{talk to GPU as 
if you were performing standard image processing}. Once again, without entering the details, 
a few steps are worth mentioning in the process:
\begin{itemize}
\item The basic computation has to be written down as a {\tt GLSL} kernel. Basic data format 
are textures on which you can copy the variables of your main program (the one which runs on the 
CPU). 
\item GLSL is initialized runtime, while your kernel has to be ``prepared'' and then
enabled to enter the rendering pipeline.
\item Input basically boils down to binding textures to texture units, while output asks for 
attaching the target texture to a {\tt FBO} (FrameBuffer Object). Finally, remember that we 
need what is called a filled quad in order to (pretend to) draw.
\item In a pipeline output of one stage is regarded as input for next one. This restricts
read/write access. A standard solution comes from the PING-PONG technique: you read from one 
texture and write onto another one. 
\end{itemize}

\subsection{Our implementation}

Also in this case, there is a natural implementation which boils down to simple recipes:
\begin{itemize}
\item Texels can be your spins and texels make a texture like spins make a lattice.
\item A {\tt RGBA} texture easily accomodates 4 independent
replicas of a lattice (maybe at different temperatures).
You only need to be careful on Metropolis acceptance
step (easily done: $Y = z Y_{\tt new} + (1-z) Y_{\tt old}$ where $z$ is a vector 
acceptance, with entries $0$ or $1$).
\item Nearest neighbors are also easily accomodated: in 2-d they are $4$, so they fit in a
{\tt RGBA} texture as well. Each spin ($4$ for each texel) knows the adrress of its own $4$ \emph{nn}, which are the {\tt RGBA} entries of a corresponding texture (of the same dimension).
\end{itemize}

Notice that with respect of such a representation 2-d systems are actually a gold plated 
application. Despite the fact that we used an old GPU ({\tt Nvidia series 6}), the speedup was 
impressive, 
provided the device was ``sufficiently filled'': 
actually, we obtained the best performance on the largest lattice we could allocate.

\section{Results}

Work was intended as a benchmark exercise.
Results are collected in Table~1 in the form of gain factors (ratios of execution times) 
with respect to the serial code (for the very same choice of parameters). 
It might well be that one can do better than this. All in all: {\tt CUDA} environment 
is really friendly and it is very easy to get a cost-effective, fairly
good performance; {\tt OpenGL} (via {\tt GLSL}) implementation was actually easier than expected and 
it delivered really good performance on a cheap device.

\begin{table}[hbt]
\begin{center}
\begin{tabular}{lll}\hline
Lattice size & Gain ({\tt CUDA}) & Gain ({\tt OpenGL} \\ \hline
128  & 33   &  1 \\
256  & 37   &  4 \\
512  & 38   & 14 \\
1024 & 30   & 41 \\
2048 & 30   & n.a. \\ \hline
\end{tabular}
\end{center}
\caption{Gain factors (ratios of execution times) with respect to the serial code.}
\end{table}

\end{document}